\documentclass[%
showpacs,preprintnumbers,
 amsmath,amssymb,
 aps,
 prl,
 longbibliography,
 lengthcheck,%
]{revtex4}

\usepackage{graphicx}
\usepackage{hyperref}

\begin{document}

\title{Random Matrix Theory and neutrino propagation in a turbulent medium}

\author{James Kneller}
\email{kneller@ipno.in2p3.fr}
\affiliation{Institut de Physique Nucl\'eaire, F-91406 Orsay cedex, CNRS/IN2P3
and University of Paris-XI, France}

\begin{abstract}
It is becoming ever clearer that the neutrino signal from the next supernova in our Galaxy 
can reveal missing information about the neutrino as well as allowing us 
to probe the explosion of the star by decoding the temporal and spectral evolution of the flavor
composition of the signal. But this information may be lost 
if turbulence in the supernova `depolarizes' the neutrinos so that the observed flux for each flavor is an equal 
mixture of the initial - unencoded - spectra. Determining if depolarization occurs is one of 
the most pressing issues of this field. The most difficult aspect of studying the effect of turbulence upon the neutrinos 
is the lack of any theoretical models that allow us to understand the results of numerical studies. This paper makes
the suggestion that Random Matrix Theory (RMT) may shine some light in this direction and presents support for this the 
possibility by comparing the distribution of crossing and survival probabilities obtained numerically for some `test case' 
calculations with the distributions one expects from RMT in the calculable limit of depolarization of N neutrino flavors. 
\end{abstract}

\date{\today}

\maketitle


\noindent
{\it Introduction}

The subject of neutrino propagation in core-collapse supernova has been the focus of intense interest in recent years
with many new effects discovered. 
From this evergrowing body of results we now can expect that 
the signal from the next Galactic supernova to reveal both unknown 
properties of the neutrinos - such as the ordering of the neutrino masses - as well as details of the explosion that would allow us 
to test the basic paradigm. This information is imprinted into the neutrino flavor composition 
of the flux as a function of time and energy due to neutrino collective effects and the evolving profile of the star as it explodes.  
For a review the reader may consult \cite{Duan:2009cd,Duan:2010bg}. 

But lurking in the shadows is the flavor transformation of the neutrino flavor due to the turbulence in the supernova.
Turbulence in a medium creates additional Mikheev-Smirnov-Wolfenstein (MSW) resonances \cite{Wolfenstein1977,M&S1986} - locations where mixing is strongest - which, together with the weaker non-resonant fluctuations, can cause flux depolarization \cite{Loreti:1995ae,Fogli:2006JCAP...06..012F,Friedland:2006ta,KV2010}. Flux depolarization is fatal for any search for information in the supernova neutrino signal because spectral features are removed or will not be imprinted. Determining if depolarization indeed occurs is a very high priority. 
But the purpose of this paper is not to reconsider the potential of the supernova neutrino signal to reveal 
properties of the supernova and/or the neutrino. Instead we grapple with the problem that studying and quantifying the effect of turbulence in supernova  lacks theoretical models that can provide insight into the problem. We make the suggestion that Random Matrix Theory (RMT) may be the doorway to progress in future studies.
 
At heart neutrino propagation through a supernova is a scattering problem in which we try to relate observed neutrino states to their initial states. 
The applicability of RMT to scattering problems has long been recognized: see, for example \cite{1997RvMP...69..731B,1998PhR...299..189G,PhysRevB.57.11258,1988AnPhy.181..290M}. That RMT may be useful for neutrino propagation is suggested by the similarity of the  non-adiabaticity parameter in Kneller \& McLaughlin \cite{Kneller:2005hf} to a Breit-Wigner resonance and recalling Ericson's prediction \cite{PhysRevLett.5.430} from RMT that of fluctuations in nuclear cross sections from intermediate states of compound nuclei. 
If RMT is indeed useful then it must be able to predict the distributions of survival and crossing probabilities obtained from numerical turbulence studies. 
This is no trivial challenge and deciding if RMT has any shot of success is left to the future. The view that RMT may be one path to progress is supported by its success in one particular case - the depolarized limit - where one can derive the expected distribution with relative ease. The plan of this paper is thus to present the results of a small numerical study of turbulence and neutrinos for some selected test cases, to derive 
the expected distributions in the depolarized limit from an ensemble of N-flavor random unitary matrices \cite{1962JMP.....3..140D}, and then make the comparison between them pointing out the similarities but also the differences that future, more sophisticated, applications of RMT will have to predict.

\noindent{\it Turbulence And Supernovae Neutrinos Test Cases}

The quantities we are interested in calculating are the survival and crossing
probabilities: i.e. $P(\nu_j\rightarrow\nu_i)\equiv P_{ij}$ since the 
flux of neutrinos emerging from the supernova is given by these quantities multiplied
by the appropriate initial fluxes at the proto-neutron star.
Throughout this paper we chose to calculate the `matter' basis probabilities
because in this basis there are no confusing adiabatic MSW transitions. The reader may find the relationship between the matter and flavor bases
in Kneller \& McLaughlin \cite{2009PhRvD..80e3002K}. These probabilities can be calculated 
from the $S$ matrix as $P_{ij}=|S_{ij}|^{2}$. The state of the neutrino after traveling a distance $r$ is related to the initial 
state at $r=0$ by $|\nu(r)\rangle = S(r,0)\,|\nu(0)\rangle$ where the matrix $S(r,0)$ is the 
solution of the equation 
\begin{equation}
\imath\frac{dS}{dr} = H\,S
\label{eq:1}
\end{equation}
with initial condition that $S(0,0)=1$. The Hamiltonian $H$ has two components: the vacuum mass term $K$ and the potential $V$ that accounts for the effect of matter. We shall not consider the case where the neutrino density is so high that collective neutrino effects need to be included. 
The vacuum mass term in the flavor basis, $K^{(f)}$, is set by the mass square differences and the parameters $\theta_{12}$, $\theta_{13}$, $\theta_{23}$ of the Maki-Nakagawa-Sakata-Pontecorvo (MNSP) mixing matrix. The two Majoranna phases of the MNSP matrix are irrelevant and we set its CP phase $\delta$ to zero.
For this paper we adopt $\delta m^2_{12}= 8 \times 10^{-5}$eV$^2$, $|\delta m^2_{23}|= 3 \times 10^{-3}\;{\rm eV}^2$, $\sin^{2} 2\theta_{12}=0.83$ and $\sin^{2} 2\theta_{23}=1$ \cite{Amsler:2008zzb}. All values of the unknown mixing angle $\theta_{13}$ we shall consider will be `large' i.e. they will be above the Dighe \& Smirnov \cite{Dighe:1999bi} threshold of $\sin^{2}(2\theta_{13})\sim 10^{-5}$. 
If the sign of $\delta m^2_{23}$ is positive then the `hierarchy' - the ordering of the masses - is known as `normal'; if the sign negative then the hierarchy is `inverted'. The potential affects only the electron neutrinos and antineutrinos and, in the flavor basis, the only non-zero element of $V^{(f)}$ is $V_{ee}^{(f)} = \sqrt{2}\,G_F\,Y_e(r)\rho(r)/m_N$ where $Y_e(r)$ is the electron fraction, and $m_N$ the nucleon mass. For simplicity we shall use $Y_e(r)=0.5$. So in order to compute the $P_{ij}$'s we require a density profile $\rho(r)$ for the neutrinos to pass through.
To study supernova turbulence $\rho(r)$ must possess fluctuations but rather than adopt density profiles taken from multi-dimensional 
hydrodynamical simulations - which presumably contain turbulence already - our approach instead is to start with a profile taken from a 
spherically symmetric simulation, i.e. a one dimensional model, that must be
turbulence free and then add turbulence. In this way we can tune the parameters that describe the turbulence plus the underlying profile
is the same for every instantiation so that non-turbulent features (such as the shocks) are always the same. 
The density profile we use is the $t=4.5\;{\rm s}$ snapshot of the $Q=3.36\times
10^{51}$ 1D hydro described in Kneller, McLaughlin \& Brockman \cite{Kneller:2007kg}. This particular simulation was chosen because it most closely
resemble the profiles of the 2D model described there. We do not distribute the turbulence throughout the entire profile because the profile is dissected
into distinct regions by the forward and reverse shocks. Within each region one would expect the turbulence, if any, to be quite different. 
For this paper we shall only insert turbulence between the reverse shock at $r_r$ and the forward shock at $r_s$ which is the region with largest amplitude fluctuations as indicated by multi-dimensional supernova simulations \cite{2006A&A...453..661K,2006A&A...457..963S,2009Nonli..22.2775G}.
The density profile $\rho(r)$ is thus constructed as $\rho(r) = \langle \rho(r)\rangle + \delta\rho(r)$ where $\langle \rho(r)\rangle$ is the adopted 
one-dimensional profile from the simulation. As is common, we model $\delta\rho(r)$ as $\delta\rho(r) = F(r)\,\langle \rho(r)\rangle$ with $F(r)$ a random field with vanishing expectation value. For this paper, $F(r)$ is proportional to: a parameter $C_{\star}$ which sets 
the amplitude of the fluctuations, the factor $\tanh((r-r_r)/\lambda)\,\tanh( (r_s-r)/\lambda)$ 
whose purpose is to avoid introducing additional discontinuities at $r_r$ and $r_s$ and 
where $\lambda$ is a scale set to $\lambda=100\;{\rm km}$, and to a Gaussian random field
described by a power spectrum $E(k)$ given by 
\begin{equation}
E(k) = (\alpha-1) \left( \frac{k_{\star}}{k} \right)^{\alpha}.
\end{equation}
The scale $k_{\star}$ is set to $\pi/(r_s -r_r)$ i.e. a wavelength twice the
distance between the shocks, and we shall use a Kolmogorov spectrum where $\alpha=5/3$. 
The `Randomization' method found in Majda \& Kramer \cite{MK1999}, using
100 wavenumbers with a cutoff at $k_{\star}$, is used to instantiate the Gaussian random field. 

\begin{figure}
\includegraphics[clip, width=\linewidth]{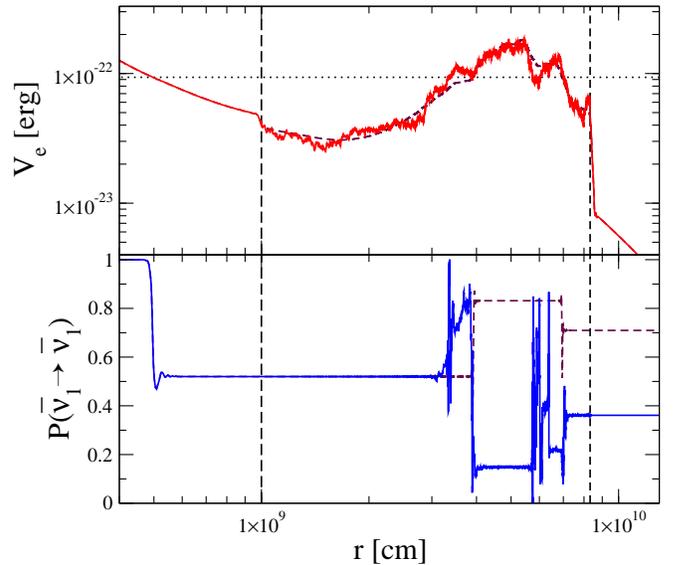}
\caption{An instance of the potential (top panel) and the survival
probability $\bar{P}_{11}$ (bottom panel) for a $E=25\;{\rm MeV}$ antineutrino as functions of the radius $r$. The turbulence region is identified
by the vertical dashed lines in both panels, the H resonance density is the
horizontal line in the upper panel. The underlying base profile from the simulation is shown as the 
dashed line, the evolution through the base profile is the dashed line in the lower panel. The fluctuation amplitude is 
$C_{\star}^{2}=0.02$, the hierarchy inverted and $\sin^{2}(2\theta_{13})=4\times10^{-4}$.\label{fig1}}
\end{figure}
Now that we have our density profiles we can calculate how a neutrino propagates through it for 
some fixed set of oscillation parameters and neutrino energy. 
The method used to solve equation (\ref{eq:1}) is described in Kneller \& McLaughlin \cite{2009PhRvD..80e3002K}. 
One finds that the probabilities $P_{ij}$ for the most past are constant as a function of the distance $r$ and change only in the vicinity of the `resonances' which are the locations where the separations of the eigenvalues of $H$ are minimal. Resonances occur at two 
different densities: the resonance at high density is known as the H resonance, the one at low density is the 
L resonance. The L resonance always occurs between states $\nu_1$ and $\nu_2$ because the sign of $\delta m^2_{12}$ is known. 
For a normal hierarchy the H resonance occurs between states $\nu_2$ and $\nu_3$: for an inverted hierarchy it is antineutrino
states states $\bar{\nu}_1$ and $\bar{\nu}_3$ that mix.
\begin{figure}
\includegraphics[clip, width=\linewidth]{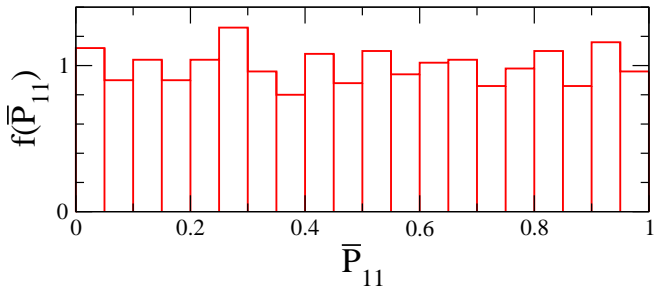}
\caption{A normalized frequency histogram of 1000 calculations of
$\bar{P}_{11}$ for a $E=25\;{\rm MeV}$ antineutrino with the underlying profile shown in figure
(\ref{fig1}), an inverted hierarchy and $\sin^{2}(2\theta_{13})=4\times10^{-4}$. \label{fig2}}
\end{figure}

\noindent
{\it Numerical Results}\\
\noindent
{\it Inverted Hierarchy}\\
With the setup complete we present numerical results; we consider first the case of an inverted hierarchy. 
In figure (\ref{fig1}) we show the evolution of the survival probability $\bar{P}_{11}=P(\bar{\nu}_1\rightarrow\bar{\nu}_1)$ as a function of
the radius $r$ through one instantiation of the potential. Without fluctuations the neutrino experiences three H
resonances; when turbulence is added the number of MSW resonances increases
dramatically. Since the value $\theta_{13}$ is `large' 
we observe that the semi-adiabaticity of the additional MSW resonances kicks
the state of the neutrino at it passes through them. But the effect of
turbulence is not just the addition of new H resonances because we also observe that non-resonant fluctuations also 
cause the probability $\bar{P}_{11}$ to evolve. Once the neutrino has passed through the entire turbulence region we find there is a
significant difference between the final states of the neutrino through the profiles with and without turbulence. This 
difference is not fixed: if we use a new instantiation of $F$ we obtain a different result. By
repeating the calculation with 1000 different instantiations we generate figure (\ref{fig2}) which is a frequency distribution 
of the final state i.e.\ the state at the edge of the star. 
The figure indicates that the distribution of the final states is consistent with
uniform and the final state is completely uncorrelated with the initial state. So for this energy, $\theta_{13}$ and turbulence amplitude we have a case where two flavor depolarization occurs.

\noindent
{\it Normal Hierarchy}

When we switch to a normal hierarchy the H resonance now affects the neutrinos. 
For modest turbulence, $C_{\star}^{2} \lesssim 0.01$, we obtain
uniform distributions of final states for $\nu_2$ and $\nu_3$. But as $C_{\star}$ increases we find something new. 
The reason is that for the normal hierarchy the L and H resonances both
occur for the neutrinos so large amplitude fluctuations break HL factorization i.e.\ the L resonances 
no longer occur after all the H resonances but rather the two are mixed together. 
\begin{figure}
\includegraphics[clip, width=\linewidth]{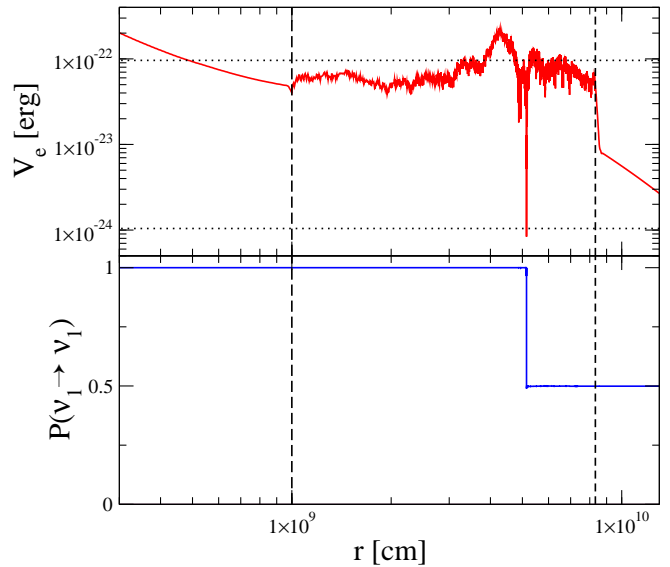}
\caption{An example of broken HL factorization. The potential (top panel) and
neutrino survival probability $P_{11}$ of a $25\;{\rm MeV}$ neutrino (bootom panel) as a function of distance $r$ 
using the same average profile shown in figure
(\ref{fig1}). The fluctuation amplitude is $C_{\star}^{2}=0.1$, the hierarchy normal and $\theta_{13}=9^{\circ}$.\label{fig3}}
\end{figure}
An example of a case with broken HL factorization can be seen in figure (\ref{fig3}). For a normal hierarchy 
the H resonance mixes states ${\nu}_2$ and ${\nu}_3$ and state ${\nu}_1$ is unaffected but for this particular calculation the density 
drops very close to the L resonance at around $r \sim 50,000\;{\rm km}$ whereupon the matter state ${\nu}_1$ mixes with matter state $\nu_2$. 
The density thereafter returns to the H resonance where further mixing between states ${\nu}_2$ and ${\nu}_3$ occurs. 
\begin{figure}
\includegraphics[clip, width=\linewidth]{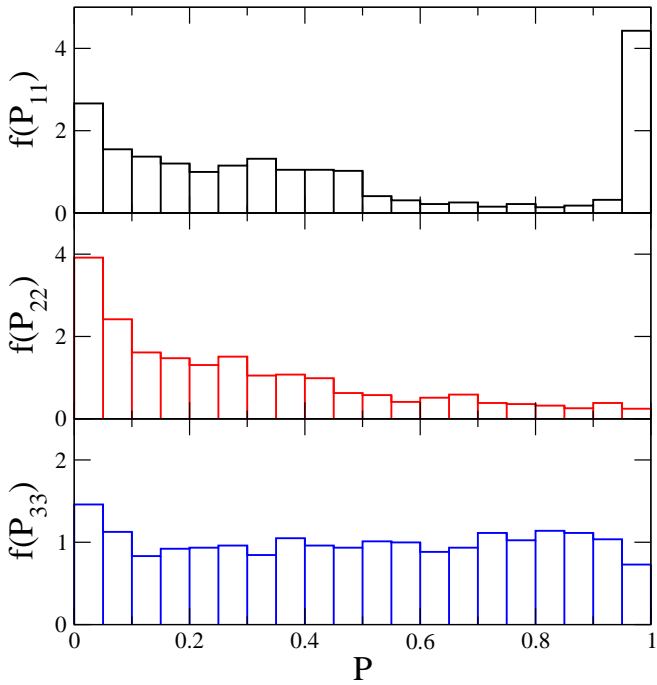}
\caption{Normalized frequency distributions of the probabilities $P_{11}$, $P_{22}$ and $P_{33}$ 
of 1563 calculations using the same average profile shown in figure
(\ref{fig1}). The hierarchy is normal, $C_{\star}^{2}=0.3$ and $E=60\;{\rm MeV}$.\label{fig4}}
\end{figure}
This mixing between states ${\nu}_1$ and ${\nu}_2$ means that the distributions for the final state probabilities are 
now quite different. The distributions of the final states for the neutrinos from 
1563 calculations for this hierarchy, $C_{\star}^{2}=0.3$ and $E=60\;{\rm MeV}$ are shown in figure (\ref{fig4}).
Without turbulence $P_{11}$ is always unity for the mixing parameters used and one notices that this is still a common result. But we also see that there
appears to be a triangular component in both the distributions of $P_{11}$ and $P_{22}$: if state ${\nu}_3$ 
has a triangular component it is not apparent. 
We thus complete our rudimentary study of turbulence effects upon supernovae neutrinos; in the next section we derive the expected distributions in the limit of N flavor depolarization and then compare them with these numerical results.

\noindent
{\it Predictions For Random Unitary $S$ Matrices}

We make the conjecture that in the depolarized limit the $S(r,0)$ matrices have a block structure such that each block 
is a random N-flavor unitary matrix \cite{1962JMP.....3..140D}. From this ansatz we can derive the distribution of the probabilities $P_{ij}=|S_{ij}|^{2}$ in this limit. 
The N real components, $x_{ij}$, plus the $N$ imaginary components, $y_{ij}$, of the elements in a row or column of the block
form a 2N Euclidean space. The requirement of unitarity defines a unit sphere in this space and since 
these 2N quantities are identically distributed the probability of a particular set of the elements must be uniform 
over the surface of the sphere. That is, the probability that we are located at $\{x_{1j},y_{1j},x_{2j},y_{2j},\ldots\}$ where, say for a column 
$S_{1j}=x_{1j}+\imath y_{1j}, S_{2j}=x_{2j}+\imath y_{2j}, \ldots$, is simply proportional to the area element $dA$ 
\begin{eqnarray}
&&P(x_{1j},y_{1j},x_{2j},y_{2j},\ldots)d^{N}xd^{N}y \propto dA \\
&&\;= \delta\left(1-\sum_{i=1}^{N}x_{ij}^{2}-\sum_{i=1}^{N}y_{ij}^{2}\right)\,\prod_{i=1}^{N} dx_{ij}\,\prod_{i=1}^{N} dy_{ij}.\;\;\;\;\label{eq:dAxy}
\end{eqnarray}
We change variables so that the $N$ independent pairs $x_{ij},y_{ij}$ are expressed as
\begin{eqnarray}
x_{1j}=\sqrt{P_{1j}}\,\cos\theta_{1j}, &\;\;\;& y_{1j}=\sqrt{P_{1j}}\,\sin\theta_{1j},\\
x_{2j}=\sqrt{P_{2j}}\,\cos\theta_{2j}, &\;\;\;& y_{2j}=\sqrt{P_{2j}}\,\sin\theta_{2j},
\end{eqnarray}
and so on. 
After inserting these new variables into equation (\ref{eq:dAxy}) and integrating over the angles we find that 
\begin{equation}
P(P_{1j},\ldots...P_{Nj})d^{N}P \propto \delta\left(1-\sum_{i=1}^{N}P_{ij}\right)\,\prod_{i=1}^{N} dP_{ij}.\label{eq:dA}
\end{equation} 
Thus, the set $\{P_{1j},\ldots P_{Nj}\}$ are uniformly distributed on the surface of a standard $N-1$ simplex. 
By integrating equation (\ref{eq:dA}) over $N-1$ of the $P$'s and normalizing one derives our final result that element $P_{ij}$ must
be distributed according to 
\begin{equation}
P(P_{ij}) =(N-1)\,(1-P_{ij})^{N-2} 
\end{equation} 
For $N=2$ the distribution is uniform with mean $1/2$ and variance $1/12$; for
$N=3$ the distribution is triangular with variance $1/18$. 

\noindent
{\it Discussion}

We find that the expected distributions in the depolarized limit possess some resemblance with the numerical results. 
For the test case results shown in figure (\ref{fig3}) we have achieved two flavor depolarization since the distribution is uniform; the results 
shown in figure (\ref{fig4}) appear to be a transitional stage between the N=2 and N=3 depolarized limits. This resemblance is encouraging and supports our suggestion that better RMT calculations - obviously not based on Dyson's ensemble - may able to predict the distributions for all cases. This suggestion is the principal message of this paper though we are fully aware that deriving the distributions for general cases will certainly be a challenge.

{\it Acknowledgements}
The authors wish to thank Cristina Volpe for the many 
thought-provoking discussions during the preparation of this article.


\end{document}